\documentclass[article]{aa} 
\usepackage{graphics,txfonts,epsf,rotating,natbib}

\bibpunct{(}{)}{;}{a}{}{,}

\def\mh{\,$\mu$Hz}
\def\hd{HD\,49933}
\begin{document}
\def\ph{\,ppm$^2$/$\mu$Hz}

\title{\Large On the detection of Lorentzian profiles in a power spectrum: \\ A Bayesian approach using ignorance priors}
\author{M. Gruberbauer\inst{1, 2}
             \and T. Kallinger\inst{1}
             \and W.W. Weiss\inst{1}
             \and D.B. Guenther\inst{2}}
             
\institute{
			Institute for Astronomy (IfA), University of Vienna,
	T\"urkenschanzstrasse 17, A-1180 Vienna, Austria \\
			\email{last~name\,@\,astro.univie.ac.at} 
			\and Department of Astronomy and Physics, Saint Marys University, Halifax, NS B3H 3C3, Canada \\
			\email{guenther@ap.stmarys.ca}}
			
\titlerunning{On the detection of Lorentzian profiles in a power spectrum}
\date{Received / Accepted }
\abstract{}
{Deriving accurate frequencies, amplitudes, and mode lifetimes from stochastically driven pulsation is challenging, more so, if one demands that realistic error estimates be given for all model fitting parameters. As has been shown by other authors, the traditional method of fitting Lorentzian profiles to the power spectrum of time-resolved photometric or spectroscopic data via the Maximum Likelihood Estimation (MLE) procedure delivers good approximations for these quantities. We, however, show that a conservative Bayesian approach allows one to treat the detection of modes with minimal assumptions (i.e., about the existence and identity of the modes).}
{We derive a conservative Bayesian treatment for the probability of Lorentzian profiles being present in a power spectrum and describe an efficient implementation that evaluates the probability density distribution of parameters by using a Markov-Chain Monte Carlo (MCMC) technique.}
{Potentially superior to ``best-fit'' procedure like MLE, which only provides formal uncertainties, our method samples and approximates the actual probability distributions for all parameters involved. Moreover, it avoids shortcomings that make the MLE treatment susceptible to the built-in assumptions of a model that is fitted to the data. This is especially relevant when analyzing solar-type pulsation in stars other than the Sun where the observations are of lower quality and can be over-interpreted. As an example, we apply our technique to CoRoT\thanks{The CoRoT ({\em COnvection, ROtation and planetary Transits} space mission, launched on 2006 December 27, was developed and is operated by the CNES, with participation of the Science Programs of ESA, ESAs RSSD, Austria, Belgium, Brazil, Germany and Spain.} observations of the solar-type pulsator HD 49933.}
{}

\maketitle
\section{Introduction}
\label{intro}
Our understanding of the Sun's structure has been revolutionized over the last three decades by the application of helioseismology, where the surface manifestation of acoustic modes (p-modes) are used to probe the interior.  Seismology is now being applied to stars using precise rapid photometry from space and high-precision radial velocity measurements from the ground. Low degree p modes were identified in several main-sequence and sub-giant stars \citep[see e.g.,][]{bed06}. 

Due to the lower signal to noise ratio of the data obtained from stars, and the more poorly constrained stellar models, there is a greater risk of getting things wrong. It is critical to a meaningful analysis to have a measure of the reliability and accuracy of the observed stellar frequency spectrum. Even a few incorrectly identified frequencies, i.e., frequencies that are not intrinsic to the star but an artifact of the data processing or of instrumental origin, will significantly complicate the analysis and may even lead to a misinterpretation of the observations. Indeed, it is a case where quality overrides quantity, that is, a few well observed modes frequencies are more useful in model fitting than a large number of poorly determined or unreliable frequencies.

Here we describe a Bayesian approach to extracting frequencies from a noisy stellar oscillation spectrum. The Bayesian method treats the problem of identifying frequency peaks in terms of probabilities. The probability that a peak in the spectrum is an oscillation mode is calculated using basic physical assumptions about the shape of the modeÕs profile and the nature of the background noise. If the data warrants it, additional prior knowledge can be included to increase the complexity of the mode shape, aiding in mode identification and the analysis of rotational effects (i.e., how frequency is split by rotation). The actual application of the Bayesian approach to compute the posterior probability distributions of models evaluated using simulated or real data is computationally complex and can quickly exceed our computational resources. Therefore, we use an implementation of the Markov-Chain Monte Carlo (MCMC) technique, which significantly reduces these demands.
The application of Bayesian treatments to solar-type oscillations is not new \citep[e.g.,][]{brewer}. The implementation of Markov-Chain Monte Carlo techniques for a Bayesian analysis of Lorentzian profiles has also already been described \citep{benomar}. 

Because we are very concerned about the reliability of the data (i.e., whether individual peaks in the spectrum are intrinsic or spurious) and specifically want to avoid interpreting noise, however, we advocate a more conservative approach. In particular, we use the mode height parameter of our models so that we can quantify the probability that a frequency peak is real or simply due to noise. This eventually allows us to arrive at a model that is just complex enough to comply with the data and its noise properties.

In the next section we describe the application of Bayes theorem to the problem of identifying stellar modes in an oscillation spectrum. In section 3 we describe our application of the MCMC technique to the computations. In section 4 we test our method on simulated data and in section 5 we apply our methodology to the recently obtained observations from CoRoT for the star HD49933. 

\section{The probability of a p-mode power spectrum model} 
In the following subsections we introduce the Bayesian formalisms including the likelihood function and the definition and role of priors. But first we review the basic properties of solar-type oscillations (e.g. \citep[see e.g.,][]{apo98}.

From the equation of a damped harmonic oscillator forced by a random function, the average power of the Fourier transform of the displacement can be approximated as,  
	\begin{equation}
	\langle P(\omega)\rangle \simeq \frac{1}{4\omega^2_0} \frac{\langle P_f(\omega)\rangle}{(\omega - \omega_0)^2 + \gamma^2},
	\end{equation}
where $P_f$ is the power of the Fourier transform of the forcing function, and $\omega_0$ and $\gamma$ are the angular mode frequency and the damping term, respectively. If the spectrum of the forcing function varies slowly with the frequency, the resulting average p-mode profile can be approximated by a Lorentzian profile as
	\begin{equation}
	\label{equ:lorprof}
	P(f, \nu, h, \tau) = \frac{h}{1 + 4(f - \nu)^2 / \eta(\tau)^2},
	\end{equation}
where $\nu$ is the mode frequency, $f$ is some frequency value in the spectrum, and $h$ and $\eta$ are the height and width ($\eta(\tau) = (\pi \tau)^{-1}$, with $\tau$ being the mode lifetime) of the profile in the power density spectrum, respectively. 
For non-radial modes $\left( \ell > 0 \right)$ stellar rotation will split the mode into a $2 \ell + 1$ multiplet. This can be modeled by adding $2 \ell$ Lorentzian profiles, spaced with integer multiples of the rotation frequency around the central mode. In this case, Equ.\,\ref{equ:lorprof} is replaced by
        \begin{equation}
	\label{equ:lorprof_rot}
	P(f, \nu, \nu_{\rm rot}, h, \tau) = \sum_{m=-\ell}^{\ell}{\frac{h_{m}}{1 + 4(f - \nu + m\nu_{\rm rot})^2 / \eta(\tau)^2}},
	\end{equation}
where $\nu_{\rm rot}$ is the rotational splitting. The mode heights $h_{m}$ of the additional profiles with $m \neq 0$ are determined by the observed geometry, specifically, the inclination angle. In general, the mode height is related to the mode amplitude according to $a^2 = \pi \eta h$. 
Usually a Maximum Likelihood Estimation (MLE) algorithm is used to find a best fit between the observed power spectrum and a model like \\ 
	$P_m(f) = b + \sum_i P(f, \nu_i, h_i, \tau_i)$,\\ 
with $b$ being the background noise (which is assumed to be locally white) and $P(f, \nu_i, h_i, \tau_i)$ being the power of the individual profiles at some frequency (-bin) 
\subsection{Application of Bayes Theorem}
The Bayes' Theorem can be stated as follows:
	\begin{equation}
	\label{equ:theorem}
	p(A | D, I) = \frac{p(A | I) p( D | A, I)}{p(D | I)},
	\end{equation}
where $p(A | D, I)$ is the probability for some hypothesis $A$ given the data $D$ and the prior information $I$. Note that $I$ is also called the posterior probability of $A$. $p(A | I)$ is the prior probability of $A$, $p(D | A, I)$ is the likelihood function, and $p(D | I)$ is the global likelihood. The latter serves as a normalizing constant and ensures the total probability summed over all hypotheses equals one. An extensive introduction to this field and the relevant methods can be found in \cite{gregory}. 

In order to apply the Bayesian formulation to any problem, one has to identify and assign probabilities for the individual terms in Equ.\,\ref{equ:theorem}. In the following subsections we discuss our choices. 

\subsection{The Likelihood Function}
\label{sec:likelihood}
The statistic of a power spectrum is given by a $\chi^2$ distribution with 2 degrees of freedom. The probability density that a model spectrum {\it m} matches an observed power spectrum {\it o} corresponds to
	\begin{equation}
	\label{equ:likelihood}
	p(m)  = \prod_{f}{\frac{1}{P_{m}(f)}e^{-({P_{o}(f)}/{P_{m}(f))}}},
	\end{equation}
where $P_{o}(f)$ is the observed power at frequency {\em f}, and $P_{m}(f)$ denotes the corresponding expectation value, i.e., the power given by the model \citep[e.g.,][]{toutain}. 
Equ.\,\ref{equ:likelihood} is used in the MLE method as a {\em likelihood function} and has already been suggested for a Bayesian treatment of solar-type oscillations \citep{appourbays, benomar}. The Bayesian formalism, though, can contain additional prior probabilities for each of the parameters from which the model spectra are constructed.

\subsection{The role of the priors - improving the MLE approach}
\label{sec:priors}
The simplest model of a p-mode profile consists of four parameters: mode frequency, mode height, mode lifetime, and background offset. Physically, the mode height and mode lifetime parameters are correlated. However, here we refrain from using this information and assume them to be independent. This enables us to use the mode height parameter as a device to distinguish between a detection and a non-detection of p-mode signal. 
We define  the following two priors, called ignorance priors because they make no assumptions about the physical properties of the object being modeled:
	\begin{enumerate}
	\item[1.] For a mode with frequency $\nu$, mode lifetime $\tau$, background offset $b$ we assume a uniform prior of the form 
	\begin{equation}
	p(x | I) = \frac{1}{x_{max} - x_{min}},
	\end{equation}  
where $x$ stands for the respective parameters $\nu$, $\tau$, and $b$. That is we assume that the probability is equal across the range of parameters and makes obvious the nomenclature, ignorance prior.  For the prior of the frequency parameter this is an appropriate choice, since we cannot {\em a priori} exclude, for an observed spectrum, any value within the range defined by the upper and lower p-mode frequencies. The mode lifetime and the background offset, as long as they do not vary across the fitted part of the spectrum by more than a magnitude, are mostly determined by the global properties of the power spectrum and therefore also warrant a uniform prior. 
\newline
\item[2.] For a mode with height $h$ we adopt a modified Jeffrey's prior of the form
	\begin{equation}
	  \label{equ:hprior}
	p(h | I) = \frac{1}{ k_h (h + h'_{min})},
	\end{equation}
where 
	\begin{equation}
	k_h = \log{\left(\frac{h'_{min} + h_{max}}{h'_{min}}\right)},
	\end{equation}
and $h'_{min}$  ($> 0$) expresses the ``strength'' of the prior. Lower values of $h'_{min}$ decrease the probability associated with $h_{max}$. In contrast to the uniform prior, the Jeffrey's prior is normally employed to ensure a uniform probability density per decade range (e.g., the prior probability between 0.001 and 0.01 is the same as between 0.01 and 0.1). The modified prior behaves like a Jeffrey's prior for $h > h'_{min}$, yet allows for values smaller than $h'_{min}$. In this regime it acts like a standard uniform prior and avoids a logarithm of 0. 
\end{enumerate}

\noindent
With the Bayesian framework restricted to ignorance priors, any evidence for significant power at some frequency $f$ can be handled by this combination of priors. 
The modified Jeffrey's prior of the mode height provides a solution for the problem that in real observations power due to noise can be mistaken for actual intrinsic signal. But by allowing the likelihood function to deliver significantly higher probabilities for mode heights well above the noise level, it reduces the mode height prior's preference for a non-detection. 
The choice of $h'_{min}$ in this context is therefore nothing more than the expression of an ``odds ratio condition''. Consider the posterior probability of the mode height at some value $h$, without considering the additional parameters, which is
	\begin{equation}
	p(h | D, I) \propto p(h | I) p(D | h, I). 
	\end{equation}
The ratio of the probabilities 
	\begin{equation}
	\label{equ:snr}
	O_{h, h'_{min}} = \frac{p(h | I)}{p(h'_{min} | I)}\frac{p(D | h, I)}{p(D | h'_{min}, I)} = O_{prior} \times O_{likelihood}
	\end{equation}
is an odds ratio for the models ``mode height\,$=h$'' and ``mode height\,$=h'_{min}$''. It can also be seen as a Bayesian weighting of a likelihood ratio, or as a strength-of-evidence indicator similar to a SNR. $O_{prior}$ is the prior odds ratio, while $O_{likelihood}$ is often called the ``Bayes factor''. 
The following equation is then an ``odds ratio condition" \emph{for the definite detection of the Lorentzian profile}:
\begin{equation}
\label{equ:oratcond}
O_{cond} = \frac{1}{O_{prior}} < O_{likelihood},
\end{equation}
If $O_{cond}$ is larger than $O_{likelihood}$, the model ``mode height\,$=h'_{min}$'' is favoured, and there is not enough evidence for the detection of a Lorentzian profile with a height greater than this threshold. For example, $O_{cond}=10^5$ requires a likelihood function value of a given $h$ that is at least $10^5$ times that of $h'_{min}$ in order for the profile to considered as detected. That this condition arises for the mode height parameter is quite intuitive, since the mode height is the parameter that determines whether a Lorentzian profile rises above the noise level. In order to ensure a constant odds ratio condition for a data set, independent of the number of modes to be detected, the {\it geometric mean} of the mode height priors can be used when more than one Lorentzian is evaluated.

To summarize, the mode height clearly is a scale parameter, rather than a location parameter \citep[see][]{gregory}. Choosing $h'_{min}$ allows one to set a lower limit for mode peak detection. The MLE procedure does not implicitly allow for such a restriction and, thus, makes it prone to overfitting. 

\subsection{Treatment of rotational effects using ignorance priors}

The effects of stellar rotation complicate the picture tremendously. If rotational splitting is suspected, one can replace single Lorentzian profiles with a sum of profiles. The central Lorentzian of the rotationally split mode can be treated just like a single profile, i.e., it has the same parameters and corresponding priors. The number of suspected split components, however, depends on the spherical degree of the mode. The components themselves are characterized by the rotational splitting and their mode heights, which relative to each other and the central peak depend on the starÕs inclination angle \citep[see, e.g.,][]{gizon} If the SNR conditions are very poor and/or a preliminary mode identification is not possible, we chose not to use the inclination angle as a parameter but to set the heights of the split components as individual free parameters (see Equ.\,\ref{equ:lorprof_rot}). Each Lorentzian profile in the model can {\it a priori} be ``equipped'' with a number of rotationally split components conforming to the highest value of $\ell$ expected to be present in the data. The corresponding height priors, then, only allow significant mode heights of those components in the rotationally split multiplets for which there is enough evidence in the data. This way, no preliminary mode identification is necessary. Alas, using this approach the number of parameters needed for the model increases tremendously. For high SNR data, where the degree of the modes becomes even visually apparent through the rotational splitting, a preliminary mode identification is certainly a more sensible approach. In such a case, though, our method is not needed anyway.

\section{Application using Markov-Chain Monte Carlo}
\label{sec:MCMC}
The analytic evaluation of complex models in a large parameter space in terms of Bayesian probability soon reaches the limit of computer resources. Stochastic methods can provide a sufficient sampling of the parameter regions of interest at a much smaller cost. In particular applications of the MCMC-technique have therefore gained increasing popularity and the (Bayesian) problems to which MCMC  nowadays are applied range from the detection of planets \citep{gregory07}, to the analysis of solar-type pulsators \citep{brewer}, to spot modelling of active stellar atmospheres \citep{croll}.  

Basically, MCMC performs a sequential walk through the parameter space of a specific problem. The MCMC technique probes solutions to the Bayesian equations by generating a random set of model parameters at some point and then progressing through parameter space via a biased random walk. This bias directing the chain of steps through parameter space is provided by the Bayesian probabilities themselves.
Specifically, each parameter of the model is incremented or decremented by some random fraction of a predefined step width and the procedure then accepts or declines this combination of steps. The condition of acceptance is provided by the {\em Metropolis-Hastings} algorithm \citep{hastings}, which is based on the ratio of probabilities of subsequent parameter configurations before and after the step. In order to comply with this algorithm, the random steps do not necessarily have to be sampled from a symmetric proposal distribution. However, using such a distribution (uniform, Gaussian, etc.) is more intuitive and facilitates the necessary calibration of the algorithm (see below). MCMC scales with an increasing number of parameters and it is an ideal tool for solving our problem. We note that the {\em Metropolis-Hastings} algorithm evaluates only the {\em relative} probability between different parameter configurations. The normalisation factor of Equ.\,\ref{equ:theorem}, which is often very difficult to evaluate, does not need to be calculated and will be dropped here after.

\subsection{Semi-automated MCMC calibration}
In order for the MCMC-implementation to operate efficiently the acceptance rate must be properly calibrated. When there are more than two parameters it can be shown that the acceptance rate should be about 0.25 in order to minimize correlations between the different parameters \citep{roberts}. Because the acceptance rate depends mainly on the step width of each parameter, the step widths, themselves, have to be calibrated. The calibration process becomes more and more cumbersome as the number of parameters increases. We, therefore, implemented the following automated MCMC calibration scheme, similar to that described in \cite{gregory}. 

Given an observed power spectrum, several non-overlapping frequency windows are defined that envelope the range of the individual mode frequency parameters. The number of Lorentzian profiles to be investigated per window is specified, and the lower and upper constraints for the mode height(s), the mode lifetime(s), etc. are defined. The model spectrum is initialized with appropriate starting parameters, such as equidistant frequencies within the frequency windows and mean values for the mode height, mode lifetime, and so on. The MCMC algorithm is started with a step width of 
	\begin{equation}
	\sigma(x) = 0.1 (x_{max} - x_{min})
	\end{equation}
for each model parameter $x$, where $x_{max}$ and $x_{min}$ are the upper and lower limits already used to define the prior probabilities.
During this ``burn-in'' phase our MCMC algorithm evaluates, for every single step and for every single parameter, the relative probability of the models according to Equ.\,\ref{equ:theorem}, using the described likelihood function and priors. It will take several hundreds to thousands of iterations to approach the parameter regions of maximum likelihood. When the ``burn-in'' phase is completed the model should be close to the global maximum of probabilities. For the next few thousand iterations, the individual acceptance rate for each parameter is evaluated every hundred or so steps. Simultaneously, the MCMC step width of the respective parameter is slowly adjusted to approach a desired acceptance rate, which depends on the total number of parameters in the model. For example, we found that in order to achieve a combined acceptance rate of about 0.25 for a model consisting of $\sim$ 70 parameters, the individual acceptance rates had to be set to roughly 0.94.

Once the desired individual acceptance rates remain fairly stable, our algorithm switches to the standard MCMC routine, where the probability of a new configuration, and therefore the probability of its acceptance, is evaluated after all parameters have been slightly changed according to a random walk. At this point the MCMC routine is set to automatically test the parameter space. 

\subsection{Evaluating the MCMC results}
Since the acceptance probability for each step is calculated using the Bayesian posterior probability, the MCMC routine will compute distributions (marginal distributions) for each involved parameter. After a sufficient number of iterations the marginal distributions for all model frequencies, mode heights, mode lifetimes, and background offsets can be estimated from histograms that count how frequently respective values of each parameter were visited. An estimate for the validity of the tested model can be obtained by examining these distributions.If any strong asymmetries, multiple local maxima, and other obviously non-Gaussian features appear they can be used to guide one toward an improved model. Regardless, the uncertainties for all parameters can be obtained from the cumulative distribution function of the normalized histograms.

\section{Simulations}
\label{sec:sim}
\begin{figure}[!ht]
	\centering
	\resizebox{ 0.9 \hsize }{!}{\includegraphics{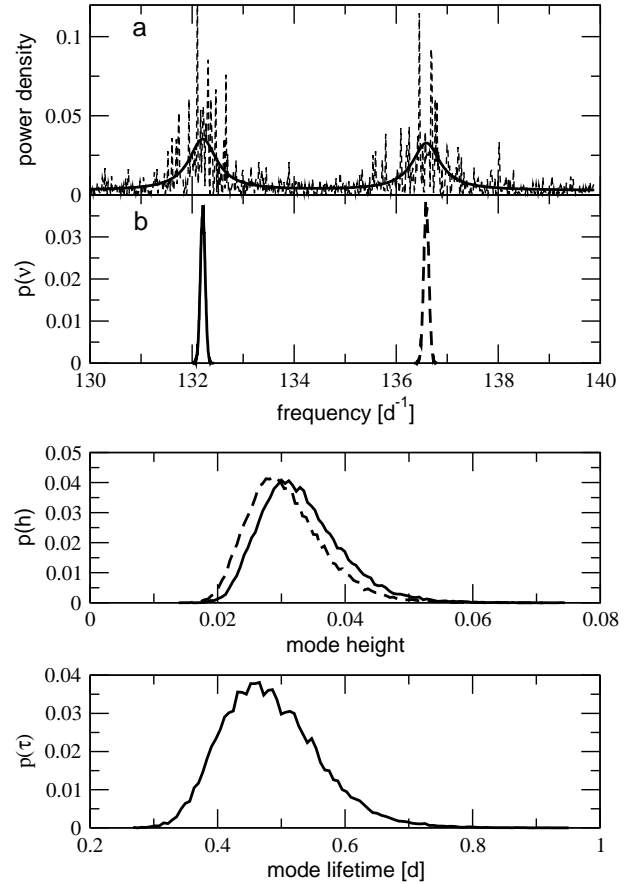}}
	\caption{Results of the Bayesian analysis for one of the 100 simulated data sets. {\em Upper panel:} (a) the power density spectrum calculated from the simulated data (dashed line) and the most probable model spectrum derived from the analysis (solid line); (b) the marginal distributions of the mode frequency parameters for both Lorentzian profiles of the model. {\em Middle panel: } the corresponding marginal distributions of the mode height parameters. Dashed and solid lines refer to the respective distributions presented in the top panel (b). {\em Bottom panel:} the marginal distribution of the mode lifetime parameter.}
	\label{fig:testspectr}
	\end{figure}

In order to test our method with parameters similar to a typical CoRoT run we generated 100 time series of 60 days duration each following the procedure in \cite{chaplin}. Each simulated data run represents a different realisation of the signal described in Tab.\,\ref{tab:sim}.
	\begin{table}[!ht]
	\caption{Input parameters for the simulated solar-type pulsator. 
	}
	\begin{center}
	\begin{tabular}{l c c}
	\hline
& $f_1$ & $f_2$ \\
	\hline
frequency & 132.3 & 136.6\\
rms amplitude & 1 & 1 \\
estimated mode height & 0.033 & 0.033 \\
mode lifetime & 0.5 d & 0.5 d \\
	\hline
\multicolumn{3}{l}{white noise model with SNR of 0.1 in the time domain } 
	\end{tabular}
	\end{center}
	\label{tab:sim}
	\end{table}
We calculated the power spectra of each time series and applied our Bayesian analysis. The frequency range was between $130$ and $140\,\rm d^{-1}$. Two Lorentzian profiles were fitted, each with adjustable mode height but equal mode lifetime. The upper and lower limits for all parameters, which also enter the Bayesian formalism via the prior probabilities, are presented in Tab.\,\ref{tab:lim}.
	\begin{table}[!ht]
	\caption{Parameter limits for the analysis of the simulated solar-type pulsator. According to Equ.\,\ref{equ:hprior}, $h'_{min}$ was set to $10^{-6}$.}
	\begin{center}
	\begin{tabular}{l c c}
	\hline
	& $min$ & $max$ \\
	\hline
	$\nu\, \rm [d^{-1}] $ & 130 & 140 \\
	$h$ &  $0$  & 0.15 \\
	$\tau\, \rm[d^{-1}]$ & 0.1 & 1.5 \\
	$b$ & 0.001 & 0.01 \\
	\hline
	\end{tabular}
	\end{center}
	\label{tab:lim}
	\end{table}
$h'_{min}$ was set to $10^{-6}$, and $h_{max}$ to 0.15, and the geometric mean of the mode height priors was used. This corresponds to an odds ratio condition of $\sim 10^{5}$ for the maximum value $h_{max}$ and to a condition of $\sim 10^{4}$ for the actual input mode heights. The results are based on 300000 iterations of the MCMC routine.

\subsection{Evaluation of the simulations}
Fig.\,\ref{fig:testspectr} shows a typical power spectrum calculated from one of the 100 test data sets, along with the most probable fit to the data. All input parameter values fall within the borders of the derived marginal distributions. We present the median values for all the parameters (but see discussion in \cite{gregory}) and, importantly, the uncertainties. A closer inspection of Fig.\,\ref{fig:testspectr} reveals that not all the derived parameter values fall within the 1$\sigma$ uncertainties of the parameter values of the simulated data. Indeed, all the parameters for this example are underestimated: The median of the $f_1$ distribution deviates from the input value by 2.99$\sigma$. The corresponding mode height is underestimated by 0.37$\sigma$. $f_2$ is underestimated by 0.63$\sigma$, with the corresponding mode height deviating by 1.2$\sigma$. The mode lifetime also deviates by 0.84$\sigma$. Such deviations are expected and are a direct consequence of the stochastic nature of the signal. 

	\begin{figure}[!ht]
	\centering
	\resizebox{\hsize }{!}{\includegraphics{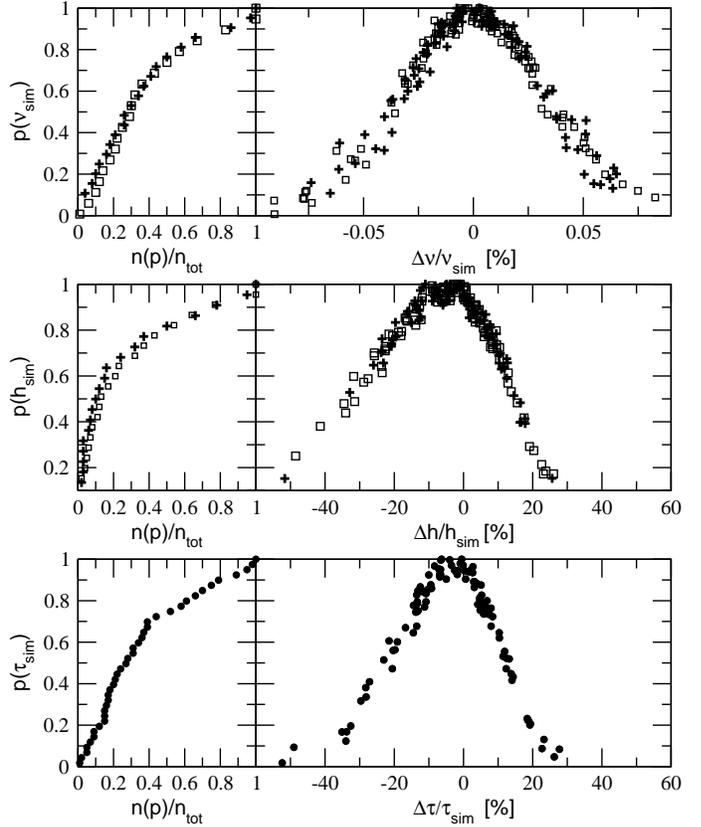}}
	\caption{{\em Right panels:} distribution of the derived median mode parameter values ($\nu, h, \tau$) compared to the input values. Positive/negative values indicate under-/overestimation of the parameters. The open squares and plus signs in the top and middle panels identify the results for the two Lorentzian profiles in the model. {\em Left panels:} cumulative histograms for the probabilities of the simulation input values determined from all 100 simulations.}
	\label{fig:allsims}
	\end{figure}

Fig.\,\ref{fig:allsims} shows the results for all 100 simulations. The median values derived from our analysis are compared to the simulation input values and are plotted against the corresponding normalized probabilities of these input values, evaluated via the marginal distributions. 
As expected, the input values are not always reproduced to within 1$\sigma$ ($p > 0.682$), but are consistently found within 3$\sigma$. The scatter of the derived median values of the parameters around the input values behaves as expected. The median of the  frequency parameter is shown to be distributed symmetrically, while the mode lifetime and mode heights follow a log-normal distribution. The figure also illustrates the overall accuracy with which each parameter can be reproduced. The frequency parameter shows a lower accuracy limit of about $0.1\%$. The corresponding lower limits for the mode height and mode lifetime parameters are much larger at about $40\%$. 

In addition, Fig.\,\ref{fig:allsims} shows cumulative histograms for the parameter value probabilities of the evaluated simulations. These histograms indicate that the scatter of the probabilities of the input values, for individual realisations, can be roughly approximated by a normal distribution. As the derived marginal distributions for all parameters provide a consistent picture, we argue that the probability densities obtained from a single data set can be trusted.

\subsection{The importance of the mode height prior}
\label{sec:exampleprior}

	\begin{figure*}[!ht]
	\centering
	\resizebox{0.65 \hsize}{!}{\includegraphics{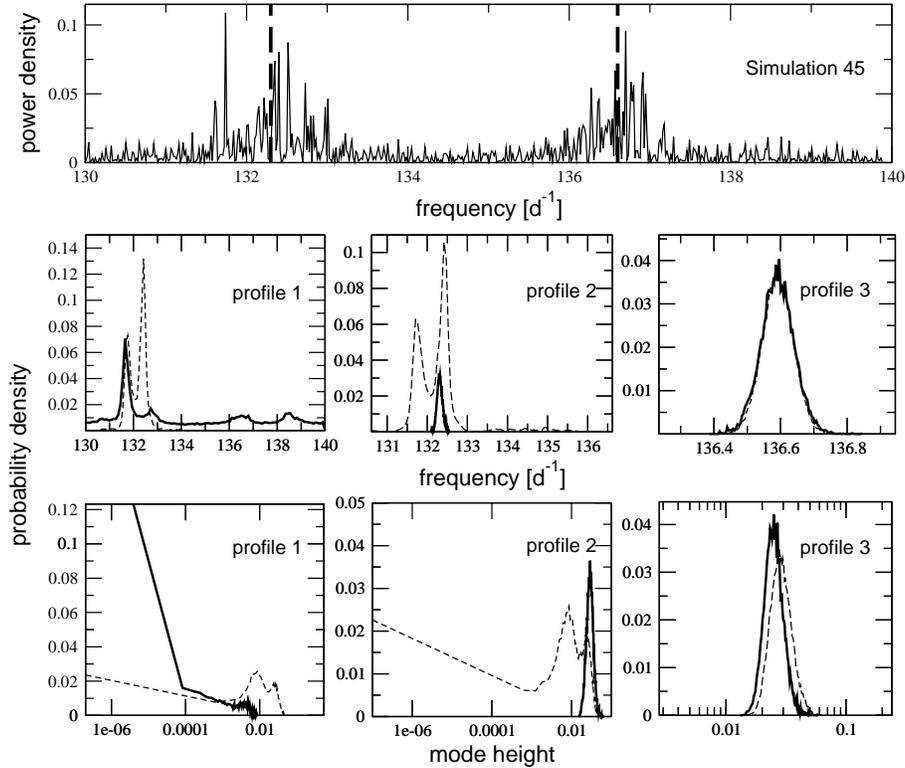}}
	\caption{3 Lorentzian profiles are fitted to simulated data that contain 2 frequencies. {\em Top panel:} power density spectrum of one of the 100 simulations mentioned in Sec.\,\ref{sec:sim}. The 2 input frequencies are marked by thick dashed lines. Additional power at $\sim 131.7 \rm d^{-1}$ is either due to noise or, more likely, due to the stochastic nature of the signal. {\em Middle panels:} marginal distributions of the frequency parameters calculated after 300000 iterations. Each panel contains the results for one of the three fitted profiles. The thick lines show the results for $h'_{min}=10^{-8}$, the thin dashed lines indicate $h'_{min}=0.01$. {\em Lower panels:} corresponding marginal distributions of the mode height parameters for each of the Lorentzian profiles and for both values of $h'_{min}$ ($10^{-8}$ and 0.01).}
	\label{fig:hminex}
	\end{figure*}

As was argued in Sec.\,\ref{sec:priors}, the mode height prior in its given form allows one to adjust the fitting to better account for noise. The simulations described in the previous Section provide us with an excellent example for this rationale. In Fig.\,\ref{fig:hminex} we show one of the simulations that includes by chance an additional power excess in the frequency range of consideration due to noise and/or the stochastic nature of the input signal. Using these data we tested how looking for of a (non-existent) third Lorentzian profile would be influenced by the $h'_{min}$ parameter of the mode height prior. We evaluated the probabilities for the existence of 3 Lorentzian profiles in the spectrum and performed the analysis with two different values of $h'_{min}=10^{-8}$ and $h'_{min}=0.01$. The weaker prior with $h'_{min}=0.01$ failed to correctly identify the noise peak as noise. This is similar to the behavior of the MLE method, or any Bayesian application that does not implicitly treat the problem of over-fitting due to, e.g., noise. The strong prior with $h'_{min}=10^{-8}$, correctly distinguished between the noise peak and the two real profiles (see Fig.\,\ref{fig:hminex}) and, in fact, determined a maximum in probability for a mode height of zero for the third. 

In the case of the \emph{strong prior}, the marginal distribution of the frequency parameters for profile 2 and profile 3 are good approximations to a normal distribution and match the two input frequencies. Indeed, profile 2 is not influenced by the additional power excess in its vicinity. Although the marginal distribution shows a maximum probability for the frequency parameter at the location of the additional power excess, the marginal distribution also assigns significant probability densities to the whole parameter range. But more to the point the marginal distributions for the height prior, which again are well-shaped for profile 2 and profile 3, show a most probable mode height of zero for profile 1. That is, the profile 1 is unnecessary to fitting this range of frequencies with only two profiles are clearly detected. 

In the case of the \emph{weak prior}, the marginal distributions of the frequency and mode height parameters of profile 1 and profile 2 intersect and mix. Hence, there seems to be evidence for a third Lorentzian profile which is comparably strong to the evidence for the actual Lorentzian profile at $132.3\,\rm d^{-1}$. Moreover, profile 1 and profile 2 cannot be easily separated. Thus, their mode parameters cannot be unambiguously determined. In summary, with a weak mode-height prior (mimicking MLE methods) a third mode is detected where none exist and the near by mode information is distorted. 

\section{Application to the CoRoT target \object{HD 49933}}

  \begin{figure}[!ht]
   \centering
      \includegraphics[width=0.5\textwidth]{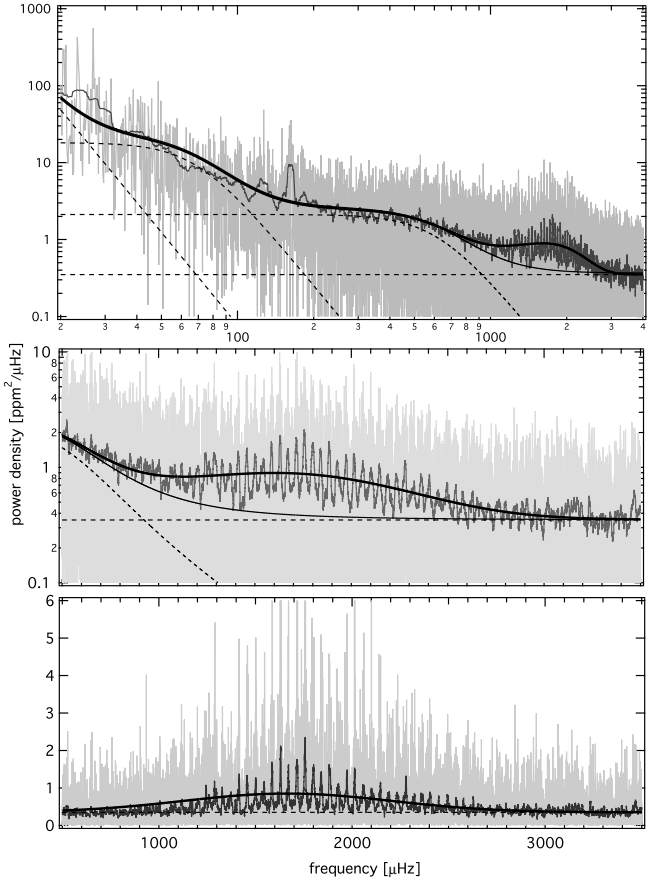}
     \caption{\emph{Top panel}: power density spectrum of HD\,49933. Light grey: original power density spectrum; dark grey: the original spectrum smoothed with a 20\mh\ boxcar average; thick solid line: global fit according to Equ.\,\ref{eq:globfit}; dotted line: global fit without the Gaussian term; dashed lines; white noise and the 3 power law components of the global fit.
      \emph{Middle panel}: enlargement of the top panel. Dotted line: background fit plus the Lorentzian profiles fitted to the residual power density spectrum.
      \emph{Bottom panel}: original power density spectrum with the granulation signal removed. Solid line: Gaussian term of the global fit.}  \label{fig:hd4spd}
   \end{figure}

The question of detecting Lorentzian profiles is only relevant for data sets, where the SNR is high enough to see evidence for solar-type pulsation, but low enough to be questionably tractable with methods used in helioseismology.
The ground-braking CoRoT data have all the qualities necessary for the technique to work:
\begin{enumerate}
\item good data quality devoid of strong instrumental signal or aliasing
\item clear indication of Lorentzian profiles at least in the region of maximum pulsation power
\item increasingly poor S/R further away from the region of maximum pulsation power.
\end{enumerate}
With the application to such data, we will demonstrate the practical utility of our Bayesian method. 
We applied our technique to the 60 days of CoRoT N2--data of \hd\ obtained during the Initial Run from February to March, 2007. A detailed description of the data extraction and reduction can be found in the \emph{CoRoT book} \citep[see][and references therein]{bag06}. The data set consists of more than 163\,000 measurements sampled each 32\,s. A detailed summary of the stellar properties and the CoRoT observations is given by \cite{apo08}.

Intrinsic stellar noise--like signal, mainly due to turbulent convection, generates significant power in the frequency range of pulsation and effects the accurate determination of mode parameters. For the Sun \cite{har85} modeled the background signal with a sum of power laws. \cite{aig04}, and more recently \cite{mic08}, use $P(\nu)=\sum_i P_i$, with $P_i=a_i \zeta_i^2 \tau_i / (1+ (2\pi \tau_i \nu)^{C_i})$, or hereafter $P_i=A_i/(1+(B_i \nu)^{C_i})$, to model the solar granulation signal, where $\nu$ is the frequency, $\tau_i$ is the characteristic times scale of the $i$th component and $C_i$ is the slope of the power law. The normalisation factor $a_i$ is chosen such that $\zeta_i^2 = \int P_i(\nu) d\nu$ corresponds to the variance of the stochastic signal in the time series. Whereas the slope of the power laws was arbitrary fixed to 2 in Harvey's original model, \cite{apo02} were the first to suggest a different value. Recently, \cite{aig04} and \cite{mic08} have shown that, at least for the Sun, the slope is closer to 4. Each power law represents a different physical process with time scales for the Sun ranging from minutes in the case of granulation to days in the case of stellar activity.
\begin{table}[t,h]
\begin{center}
\caption{Global fit parameters. The amplitudes of the power laws (A$_i$), the height of the Gaussian part (P$_g$), and the white noise components (P$_n$) are given in \ph . The power law time scales (B$_i$) are given in seconds and the center ($\nu_\mathrm{max}$) and width ($\sigma$) of the Gaussian part are in \mh . One-sigma error estimates are given in brackets.}
\label{tab:globfit}
\begin{tabular}{lllll}
\noalign{\smallskip}
\hline
\hline
\noalign{\smallskip}
             	      && A$_1$ &  A$_2$ &  A$_3$ \\
                      &  & B$_1$ &  B$_2$ &  B$_3$ \\
\noalign{\smallskip}
\hline
\noalign{\smallskip}
Power laws	&&25743 (6200) 	& 22 (6)	& 2.2 (0.11)	\\
	 		&&252311 (25000) & 16477 (1950)	& 1637 (62)	\\
\noalign{\smallskip}
\hline
		&& P$_g$	& $\nu_\mathrm{max}$	& $\sigma$\\
\hline
\noalign{\smallskip}
Gaussian term && 0.500 (0.015) & 1657 (28) & 538 (26) \\
\noalign{\smallskip}
\hline
White noise	&& \multicolumn{3}{c}{P$_n$ = 0.33 (0.005)}\\
\hline
\noalign{\smallskip}
\end{tabular}
\end{center}

\end{table}

\cite{apo08} used a single power law for the background signal in the frequency range of pulsation and fitted the corresponding parameters simultaneously with the $p$-mode parameters to the observed power density spectrum. 
We have separated the analysis of the background signal from the analysis of the pulsation signal, in which case one has to model the pulsation signal in the power density spectrum with a dedicated term in the global background fit.
\cite{kal08} have shown that the power excess hump due to pulsation can be approximated by a Gaussian function. Hence, our global model to fit the heavily smoothed power density spectrum of \hd\ consists of a superposition of white noise, three power law components, and a power excess hump approximated by a Gaussian function:
	\begin{equation}
	P(\nu) =  P_n +  \sum_{i=1}^{3}\frac{A_i}{1 + (B_i  \nu)^4} + P_g \cdot e^{-(\nu - \nu_{max})^2/(2\sigma^2)},
	\label{eq:globfit}
	\end{equation}
where $P_n$ is the white noise component, $A_i$ and $B_i$ are the amplitudes and characteristic time scales of the power laws, $P_g$, $\nu_\mathrm{max}$, and $\sigma$ are the height, central frequency, and width, respectively, of the Gaussian term. The resulting global fit (and its components) is shown in the top and middle panel of Fig.\,\ref{fig:hd4spd} along with the original and heavily smoothed power density spectrum. The corresponding parameters of the global fit are given in Tab.\,\ref{tab:globfit}. The fit without the Gaussian component (dotted line in top and middle panel of Fig.\,\ref{fig:hd4spd}) enables us to estimate the local noise and is used to separate the background signal form the pulsation signal. The residual power density spectrum rescaled to the white noise level, shown in the bottom panel of Fig.\,\ref{fig:hd4spd}, is used for the subsequent analysis.

\subsection{MCMC analysis}

The frequency range between the lower and upper limits beyond which we see the pulsation signal disappearing in the noise was subdivided into 16 windows. To each, a chosen number of Lorentzian profiles was fitted, based on our Bayesian algorithm. The model parameters are presented in Tab.\,\ref{tab:corlim}. This subdivision into windows was only chosen to accelerate the burn-in phase. It has no influence on the results, as long as the resulting marginal distributions for the mode frequency parameters are not intersected by the window borders. Yet, it still enables us to perform a global fit which includes the influence of Lorentzian profiles also from adjacent windows.
	\begin{table}[!h]
	\caption{Parameter limits for the analysis of HD49933. The height prior parameter was set to $h'_{min}=10^{-8}$.}
	\begin{center}
	\begin{tabular}{l c c}
	\hline
& $min$ & $max$ \\
	\hline
	\hline
$\nu\, \rm [\mu Hz] $ & 1219.9 & 2618.1 \\
$h\, [ppm^{2}/\mu Hz]$ &  0  & 6 \\
$\tau\, \rm[d^{-1}]$ & 0 & 1.5 \\
$b\, [ppm^{2}/\mu Hz]$ & 0.001 & 0.5 \\
	\hline
	\end{tabular}
	\end{center}
	\label{tab:corlim}
	\end{table}
	
For the first analysis we decided to consider only 2 Lorentzian profiles in each window. Moreover, we also kept the mode lifetime uniform for all Lorentzian profiles, expecting that the marginal distribution of this parameter would tell us whether this assumption is valid. 

	\begin{figure*}[!ht]
   	\centering
      \includegraphics[width=0.95\textwidth]{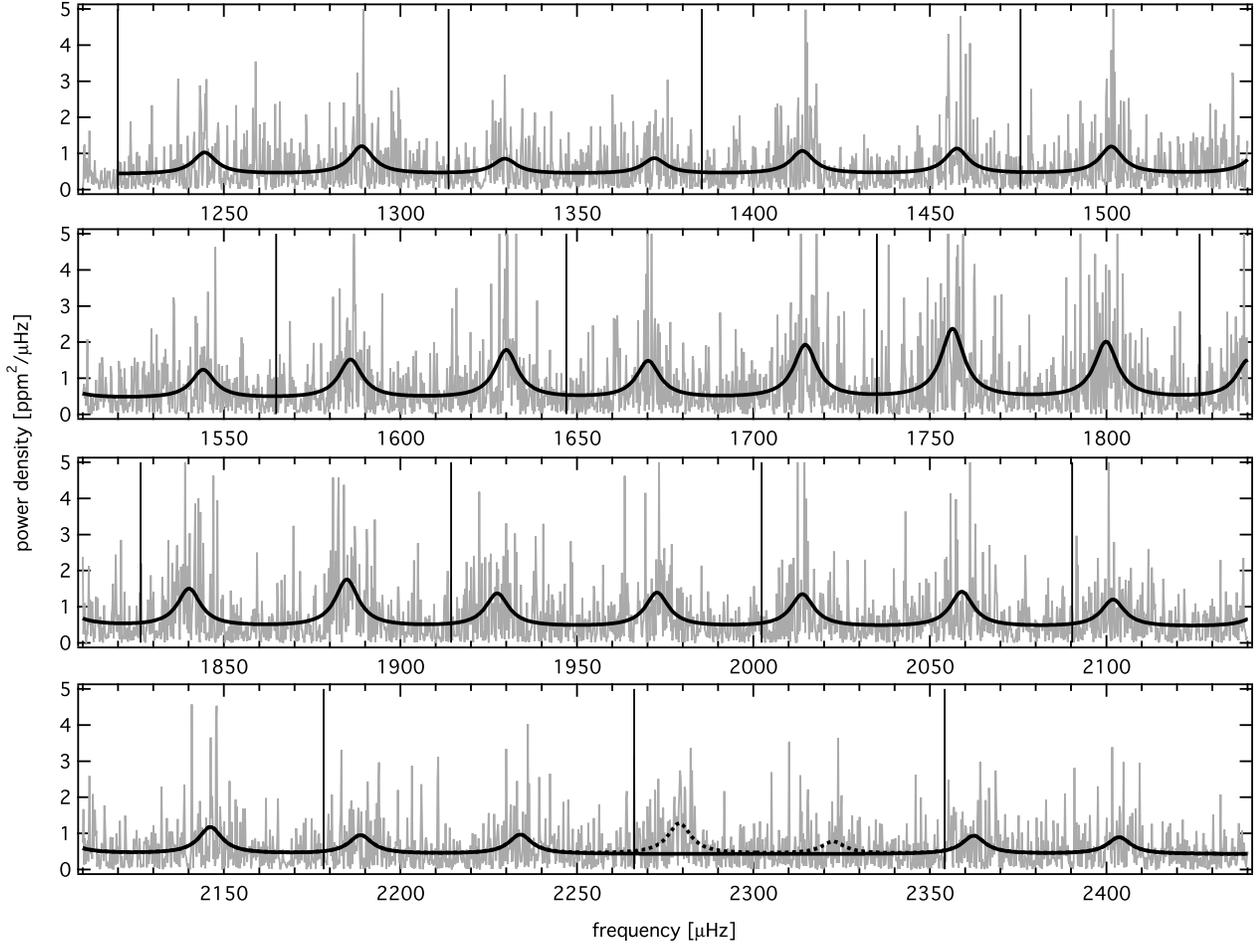}
      \caption{Observed power density spectrum (gray) and fitted Lorentzian profiles.Vertical lines indicate the borders of the fitting windows (see text). The dashed lines in the bottom panel show two profiles for which our Bayesian technique results in ambiguous frequency distributions ranging over the whole fitting window}
   	\label{fig:specfit}
   	\end{figure*}
   
Fig.\,\ref{fig:specfit} shows the power density spectrum, on which our analysis is based, together with the fitting windows we defined, and the most probable model derived after $\sim 1.3$ million iterations. The model manages to reproduce the observations quite well, and there remain only very few frequency ranges where a visual inspection seems to indicate additional power.
Beyond $\nu > 2450\, \rm \mu Hz$ several Lorentzian profiles are fitted with fairly well defined frequencies, but with a probability maximum for a mode height of zero. Two profiles, listed in Tab.\,\ref{tab:corresults} as P25 and P26, indicated in the figure by dashed lines, have ambiguous frequency distributions ranging over the whole fitting window. What is shown is a biased choice to fit the general picture. 

	 \begin{figure}[!ht]
   	\centering
     	\includegraphics[width=0.4\textwidth]{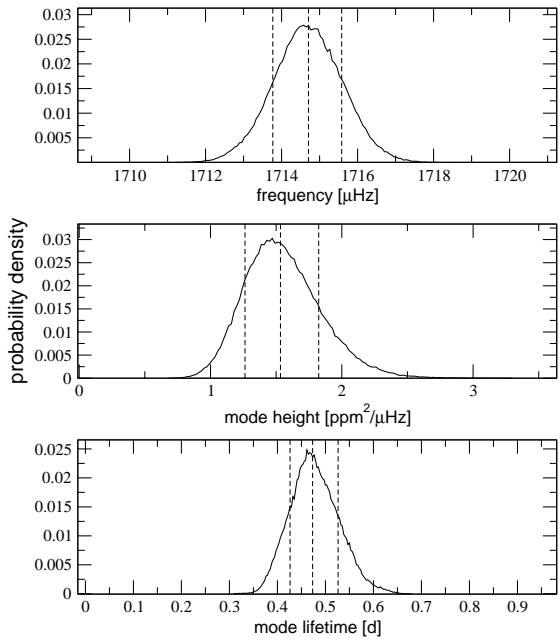}
      	\caption{\emph{Top panel}: marginal distribution of the mode frequency parameter of \textbf{one} of the profiles fitted to the CoRoT data. The median value, and the lower and upper 1$\sigma$-values are indicated by the dashed lines. \emph{Middle panel}: same as top panel but for the corresponding mode height parameter. \emph{Bottom panel}: same as top panel but for the mode lifetime parameter for a simultaneous fit to \textbf{all} Lorentzian profiles.}
  	 \label{fig:resexample}
   	\end{figure}
Fig.\,\ref{fig:resexample} presents an example for the marginal frequency distribution and mode height for one of the Lorentzian profiles. As expected, the former can be very well approximated by a Gaussian, while the latter follows a log-normal distribution. The bottom panel shows a quite narrow and smooth mode height distribution for a simultaneous fit to all Lorentzian profiles, indicating a constant mode lifetime. The data therefore do not seem to warrant different mode lifetimes. It is thus save to assume that any variations among the various profiles for this parameter are accounted for within the uncertainty given by the mode lifetime distribution. 
Surprisingly, we do not find well-defined multi-modal distributions for any of the frequency parameters involved. Some ambiguities in the distributions are only apparent at very high frequencies and with the poorest SNR. Additional profiles, either due to modes of higher degree or rotational splitting, still could be present, if the frequencies are very well separated.
  	\begin{figure}[!ht]
  	\centering
     	\includegraphics[width=0.45\textwidth]{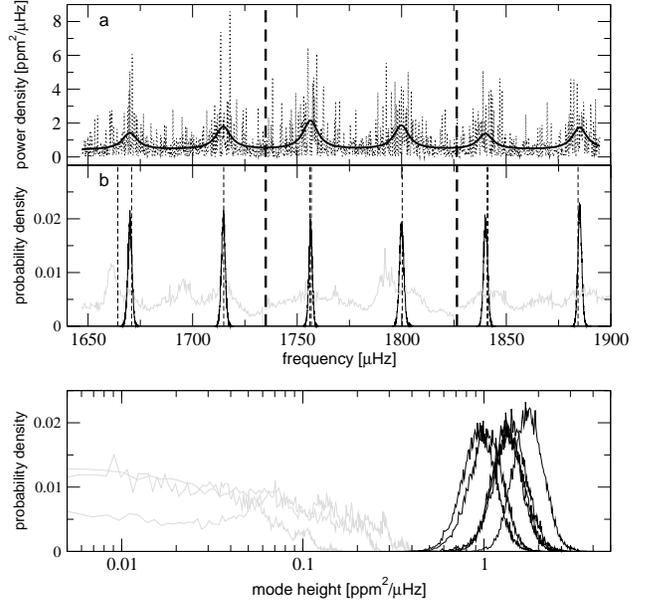}
     	\caption{\emph{Top panel:} (a) the fitted region in the power density spectrum, where the p modes have the highest SNR. The most probable model (thick solid line) is presented in comparison to the observations (dotted line). The borders of the fitting windows are indicated by the thick dashed lines. (b) The marginal mode frequency distributions for 3 profiles per window. Six profiles are identified (black), while the additional profiles are ambiguous across the whole frequency range (gray lines). Frequencies identified by \cite{apo08} are shown as thin dashed lines. \emph{Bottom panel:} The corresponding mode height distributions.}
   \label{fig:3rdmode}
   \end{figure}

\subsection{On the evidence for $\ell=2$ modes}

We repeated our analysis for the region with the highest SNR, but included for each fitting window a third Lorentzian profile in our model. The results are shown in presented in Fig.\,\ref{fig:3rdmode}. The probability distributions for the additional profiles consistently favours a mode height of zero. Moreover, the corresponding mode frequency distributions only vaguely contain regions of higher probability. We therefore conclude that, based on our conservative approach, any additional power in the power density spectrum is due to noise or to the stochastic nature of the clearly detected modes. From our perspective, focussing on the ``detection'' of profiles, the data do not present convincing evidence for $\ell=2$ modes.  

\subsection{On the evidence for rotational splitting}
\label{sec:evirot}
\begin{figure}
   \centering
      \includegraphics[width=0.45\textwidth]{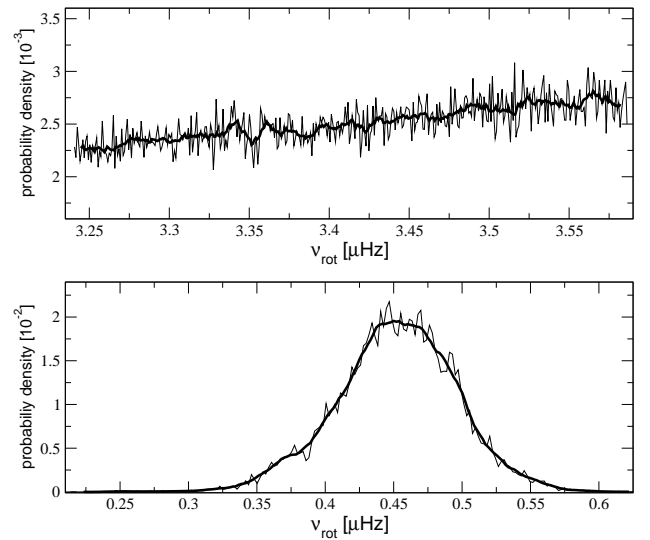}
      \caption{\emph{Top panel}: the marginal distribution of the rotational splitting parameter derived as explained in Sec.\,\ref{sec:evirot}. The thick black line is a boxcar average using 10 data points. \emph{Bottom panel}: the same but for the analysis of VIRGO data of the Sun.}
   \label{fig:rot}
   \end{figure}

The failed detection of additional profiles, and the single-mode nature of the frequency distributions of detected profiles, suggests that the effects expected from rotational splitting can most likely be neglected in the analysis of the CoRoT data of \hd. Nonetheless, in order to perform a more rigourous test, we analysed the same frequency region shown in Fig.\,\ref{fig:3rdmode} including these effects. Our model used 2 Lorentzian profiles per fitting window, each of which contains 2 additional features corresponding to rotational splitting of $\ell=1$ modes (see Equ.\,\ref{equ:lorprof_rot}). Accordingly, a new global parameter, the rotation frequency, was introduced. \cite{apo08} report the detection of a particular spectral feature at 3.4\,\mh\ which apparently corresponds to the \hd's rotation frequency. We therefore used this value as a reference and allowed a scatter of about 0.2\,\mh\ in either direction. This range corresponds to twice the classical frequency resolution $(\Delta T)^{-1}$ of the CoRoT time series, where $\Delta T$ is the length of the data set, and was used to define the borders of a uniform prior. The heights of the rotation profiles were implemented as usual, using modified Jeffrey's priors for these parameters. 
We expected a well defined distribution for the rotation frequency to emerge from the analysis. However, we find no preferred value for this parameter, as is shown in Fig.\,\ref{fig:rot}. The distribution is mostly consistent with sampling noise. 

Concerning the mode heights, alternating detections and non-detections of the rotation profiles would indicate a difference in their central profiles between radial and non-radial modes. We cannot find any such differences, since the mode height distributions for the rotation profiles are consistent with a null result. While the star certainly does rotate, it seems that the signal is too close to the noise level to support any need for these parameters. In other words, rotational effects, if present, seem to be ill--defined and do not need to be considered for the frequency analysis of this data set. As an example of a near ideal case, the bottom panel of Fig.\,\ref{fig:rot} shows corresponding results from a single mode using 60 days of data of the Sun observed as a star by the VIRGO (green band) instrument on board the SOHO spacecraft \citep[see][and references therein]{froehlich97b}. Fitting only the small range between 3090 and 3110 \mh\, and using the same profile model as for \hd, the rotational splitting becomes apparent.

\subsection{Results}

\begin{figure}
   \centering
      \includegraphics[width=0.4\textwidth]{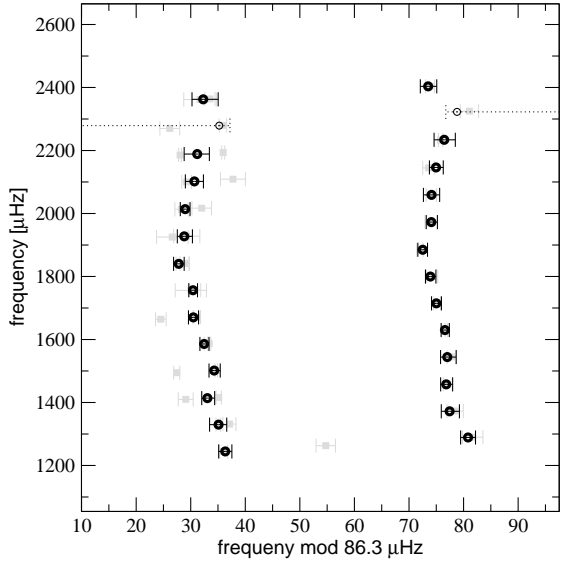}
      \caption{Echelle diagram showing the identified frequencies (solid error bars). Values belonging to profile 25 and 26 (see Tab.\,\ref{tab:corresults}) are indicated by the dashed error bars. Frequencies found by \cite{apo08} are displayed in grey.}
   \label{fig:echelle}
   \end{figure}

As a consequence of the results from the previous two sections, we use only 2 modes per window without any rotational splitting for our global fit. For 6 of the 32 fitted profiles, the identification is ambiguous (or the most probable mode heights are close to zero) due to the low SNR. All other profiles are clearly detected and their parameters have smooth, single-mode distributions. The results of our analysis are presented in Tab.\,\ref{tab:corresults}. The 1$\sigma$-errors have been derived from the cumulative distribution functions, which are evaluated using the marginal distributions for all parameters. We find that the data are consistent with a mode lifetime of roughly 0.5 days which does not appear to vary significantly across the spectrum. The confidence for this overall result can be calculated via the mode height parameter. As we have used $h'_{min}=10^{-8}$, the result gives an odds ratio condition $O_{cond} = 4.07 \times 10^{6}$. 
\newline
Therefore, the  obtained parameter distributions are at least $4.07 \times 10^{6}$ times more probable than a model spectrum that only consists of the background offset. The profiles 25 and 26 are only listed for sake of completeness. Due to the ambiguity in their marginal frequency distribution, apparent in their given uncertainties, we do not assign credibility to these values. The final Echelle diagram is displayed in Fig.\,\ref{fig:echelle}. The frequencies estimated as $\ell=1$ by \cite{apo08} agree very well with our corresponding values, and the 1$\sigma$-uncertainties are comparable in both studies. The remaining frequencies, however, differ due to the cited authors' explicit assumption of $\ell=2$ modes, which we do not detect. 

\begin{figure}
   \centering
      \includegraphics[width=0.5\textwidth]{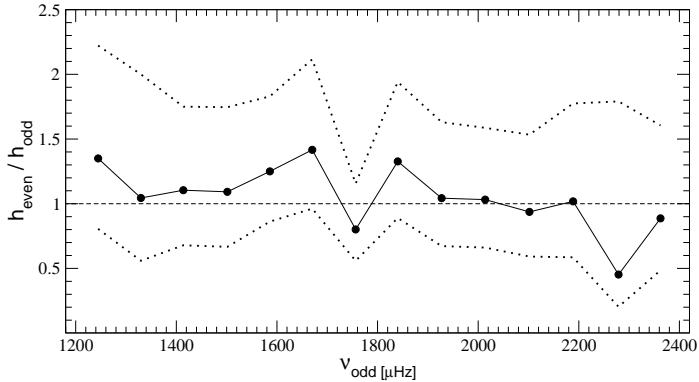}
      \caption{The mode height ratios for subsequent modes listed in Tab.\,\ref{tab:corresults} (e.g., $h_{P2}/h_{P1}$) are shown as filled circles. The dotted lines correspond to the $68.2 \%$ confidence limits. The dashed line indicates a height ratio of unity. The outlier at $\sim 2280\, \rm \mu Hz$ is due to profiles 25 and 26, which are ambiguous detections.}
   \label{fig:amprat}
   \end{figure}

Although we deliberately did not include fixed mode height ratios for modes of different degree in our model, the resulting mode height ratios are mostly consistent within the uncertainties with a fixed ratio. The obvious outlier in Fig.\,\ref{fig:amprat} at $\sim 2280\, \rm \mu Hz$ is due to profiles 25 and 26, which we already marked as highly suspicious.
\begin{table}
\caption{Results of the CoRoT data for HD 49933. Profiles P 25 and P 26 are ambiguous detections. Profiles P 29 to P 32, which 
are not listed, have a most probable mode height of zero.} 
\begin{tabular}{l l c l c}
\noalign{\smallskip}
\hline
\hline
\noalign{\smallskip}
P & $\nu$ &  $\sigma_\nu$ & h & $\sigma_h$ \\  
& [$\mu$Hz] & & [ppm$^2$/$\mu$Hz]  & \\
\noalign{\smallskip}
\hline
\noalign{\smallskip}
1 &   1244.53	& (-1.21 / +1.21) & 0.60  & (-0.15 / +0.17)  \\
2 &   1289.02	& (-1.36 / +1.36) & 0.81  & (-0.19 / +0.19)  \\
3  &   1329.60	& (-1.50 / +1.64) & 0.45  & (-0.14 / +0.14)  \\
4  &   1371.94	& (-1.80 / +1.53) & 0.47  & (-0.14 / +0.15)  \\
5  &   1413.85	& (-1.34 / +1.03) & 0.67  & (-0.15 / +0.17)  \\
6  &   1457.63	& (-1.17 / +1.02) & 0.74  & (-0.17 / +0.17)  \\
7  &   1501.40	& (-1.11 / +0.95) & 0.76  & -0.17 / +0.20)   \\
8  &   1544.11	& (-1.63 / +1.22) & 0.83  & (-0.19  / +0.20) \\
9  &   1585.86	& (-0.90 / +0.78) & 1.08  & (-0.20 / +0.22)  \\
10  &   1629.99	& (-0.81 / +0.70) & 1.35  & (-0.23 / +0.26)  \\
11  &   1670.18	& (-0.96 / +0.89) & 1.08  & (-0.22 / +0.23)  \\ 
12  &   1714.71	& (-0.94 / +0.88) & 1.53  & (-0.27 / +0.29)  \\
13  &   1756.42	& (-0.82 / +0.76) & 2.01  & (-0.34 / +0.37)  \\
 14 &   1799.91	& (-1.02 / +0.89) & 1.61  & (-0.28 / +0.32)  \\
15  &   1840.10	& (-1.00 / +0.94) & 1.04  & (-0.19 / +0.23)  \\
16  &   1884.81	& (-0.89 / +0.89) & 1.38  & (-0.25 / +0.27)  \\
17  &   1927.41	& (-1.50 / +1.27) & 0.93  & (-0.20 / +0.23)  \\
18  &   1972.70	& (-1.10 / +0.94) & 0.97  & (-0.19 / +0.22)  \\
19  &   2013.88	& (-0.89 / +0.89) & 0.94  & (-0.19 / +0.21)  \\
20   & 2059.04 &  (-1.49 / +1.49) &0.97  & (-0.21 / +0.22)  \\
21  &   2101.88	& (-1.66 / +1.66) & 0.79  & (-0.19 / +0.19)  \\
22  &   2146.16	& (-1.32 / +1.22) & 0.74  & (-0.16 / +0.18)  \\
23  &   2188.69	& (-2.23 / +2.39) & 0.55  & (-0.15 / +0.15)  \\
24  &   2233.96	& (-2.03 / +1.87) & 0.56  & (-0.15 / +0.15)  \\
{\em 25} &  {\em 2279.03} &  {\em (-1.96 / +41.67)} & {\em 0.84} & \em (-0.41 / +0.24) \\
{\em 26} &  {\em 2322.60} &  {\em (-44.02 / +2.06)}  & {\em 0.38} & {\em (-0.16 / +0.39)}  \\
27  &   2362.40	& (-2.74 / +2.06) &  0.50  & (-0.15 / +0.15)  \\
28  &   2403.61	& (-1.58 / +1.43) &  0.46  & (-0.14 / +0.14) \\
\noalign{\smallskip}
\hline
\hline
\noalign{\smallskip}
 & $\tau$ & $\sigma_\tau$ & $\eta$&  $\sigma_\eta$ \\
 &  [d] & &  [$\mu$Hz] & \\
 \noalign{\smallskip}
 \hline
 \noalign{\smallskip}
all & 0.47 & (-0.04 / + 0.06) & 7.8 & (-0.8 / +0.8) \\
\noalign{\smallskip}
 \hline
 \hline
 \noalign{\smallskip}
& b & $\sigma_b$ & & \\
& [ppm$^2$/$\mu$Hz] & & & \\
\noalign{\smallskip}
 \hline
 \noalign{\smallskip}
all & $0.42$ & $(-0.02 / +0.02)$ & & \\
\noalign{\smallskip}
\hline
\end{tabular}
\label{tab:corresults}
\end{table}

\section{Discussion}
With high quality data and evidence for additional and/or closely spaced modes, such as might appear for nonradial modes split by rotation, we believe that it might be advantageous to change the peak Lorentzian model by combining the corresponding Lorentzian profiles and fitting them as a group. Eventually, once the quality of the data becomes even better, we think that a more classical Bayesian treatment replacing the ignorance priors with specific information from theory or previous observations, is more appropriate because the potential gain in information outweighs the danger of over-fitting. \cite{benomar} has performed such an analysis for the data set also presented in this paper, and obviously obtained different results. It thus remains questionable whether the Initial Run CoRoT data of HD 49933 is already good enough for such an approach.

\subsection{On the global likelihood of p-mode profiles}
In this paper we have used Bayes' theorem to solve a parameter estimation problem. One of the biggest advantages of Bayesian analysis, however, is to perform model comparison. This is achieved by first calculating the global likelihood via integration over all model parameters. The global likelihoods are subsequently used to compare the different models themselves. It would be interesting to use parallel tempering \citep[e.g.,][]{benomar}, an algorithm which allows the Markov chain to more easily sample the complete parameter space, to perform such model selection by calculating the global likelihood of a model via integration over the tempering parameter \citep[see, e.g.,][]{gregory07}. However, we think that the validity of applying this kind of model selection to p-mode analysis needs to be carefully tested first. 

\begin{figure}
   \centering
      \includegraphics[width=0.5\textwidth]{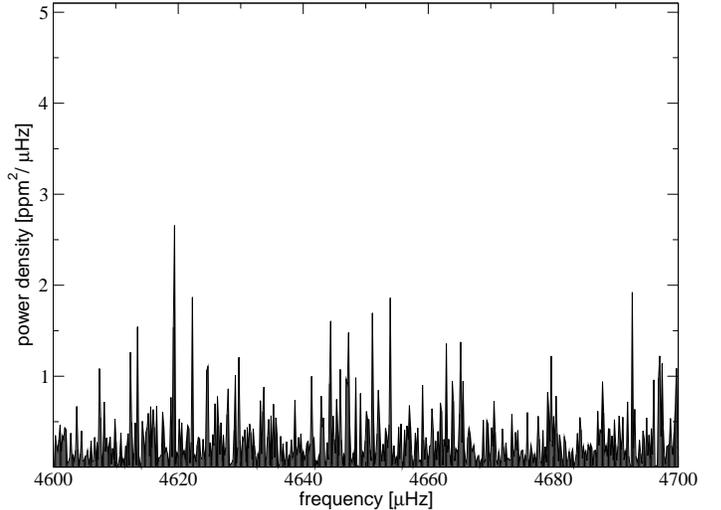}
      \caption{The high-frequency region devoid of p-mode signal used for testing the global likelihood comparison. Note that the scale is similar to Fig.\,\ref{fig:specfit}.}
   \label{fig:noisetest}
   \end{figure}

As a first trial run, we have investigated how the global likelihood of three simple models behaves in a region of pure noise. We studied the region between 4600 and 4700 \mh\, in the power spectrum of HD 49933 (see Fig.\,\ref{fig:noisetest}) which is far beyond the region of pulsation. At such high frequencies, the power spectrum should be dominated by the noise properties of the instrument. Following the description given in \cite{gregory} on how to obtain the global likelihood, we evaluated three models
\begin{itemize}
\item M1: constant noise level,
\item M2: constant noise level + 1 Lorentzian profile, and
\item M3: constant noise level + 2 Lorentzian profiles 
\end{itemize}
The noise level was allowed to vary between 0.001 and 0.5 $\rm ppm^2$\mh$^{-1}$. All Lorentzian profiles were allowed a mode lifetime between 0 and 1.5 days, mode heights between 0 and 6 $\rm ppm^2$\mh$^{-1}$, and frequencies between 4600 and 4700\mh. We used 10 chains with decreasing but equidistant tempering parameter. In order to study the effects of the likelihood function on the gobal likelihood we neglected all priors in the calculations. The results are best expressed using the odds ratios  
\begin{eqnarray*}
\frac{p(M1 | D, I)}{p(M2 | D, I)} \approx exp(-240) & \rm and & \frac{p(M2 | D, I)}{p(M3 | D, I)} \approx exp(-27). 
\end{eqnarray*}	
Depending on the choice of the tempering parameters used in the parallel tempering process, as well as on differences in the numerical integration scheme required to calculate these numbers, the end result varies, but the overall magnitudes of difference in probability remain the same. The likelihood function alone seems to prefer models with (more) Lorentzian profiles, even if there are none (related to solar-type oscillations) present in the data. 

This is an issue demanding clarification. It will be required to test the influence of the allowed parameter ranges in order to see where this preference arises in the integration over the parameter space. Moreover, the influence from different kind of prior probabilities on the global likelihoods also needs to be studied. To conclude, our initial test suggests that we cannot yet trust the global likelihood completely when comparing models of power spectra (to real data at least) until we are able to clarify this issue. 

\section{Conclusion}
We have presented a conservative approach (using only ignorance priors) for a Bayesian analysis of solar-type p modes with a minimum of parameters. The method uses a mode-height prior that allows us to more easily and reliably distinguish real peaks from those that arise stochastically from noise, in reference to a threshold set by the user. We show how the marginal distributions of all the parameters can be obtained. Our tests with simulated data succeeded by reproducing the simulation input values and not being confused by randomly occurring noise peaks. 
Furthermore, we have applied our method to the Initial Run CoRoT data of HD 49933, where we can easily and unambiguously identify at least 26 p modes. We fail to detect $\ell=2$ modes or effects produced by rotation. We also approached the possibility of performing Bayesian model comparison in the analysis of solar-type p modes. First results suggest, however, that more tests need to be performed in order to understand how model comparison works in a power spectrum.
Finally, we would like to point out a companion paper \citep{kall_neu}. Therein, we show that frequencies extracted in this work are in excellent agreement with model frequencies of a solar-calibrated model that coincides with the spectroscopically determined position of \hd\ in the H-R diagram.

 \begin{acknowledgements}
We would like to especially thank Daniel Huber (Sydney Institute for Astronomy, Sydney, Australia) for interesting discussions. We also thank Peter Reegen (Institute for Astronomy, Vienna, Austria) and Tim Bedding (Sydney Institute for Astronomy, Sydney, Australia) for their suggestions. MG, TK and WW have received financial support by the Austrian Fonds zur F\"orderung der wissenschaftlichen Forschung (P17890-N02). DBG acknowledges funding from the Natural Sciences \& Engineering Research Council (NSERC) Canada. Last but not least, we are very thankful for the constructive comments provided by the anonymous referee who helped to greatly improve this article. We are grateful for the VIRGO data being publicly available.
\end{acknowledgements}
\bibliographystyle{aa}
\bibliography{11203}

\begin{thebibliography}{22}
\expandafter\ifx\csname natexlab\endcsname\relax\def\natexlab#1{#1}\fi

\bibitem[{{Aigrain} {et~al.}(2004){Aigrain}, {Favata}, \& {Gilmore}}]{aig04}
{Aigrain}, S., {Favata}, F., \& {Gilmore}, G. 2004, A\&A, 414, 1139

\bibitem[{{Appourchaux}(2008)}]{appourbays}
{Appourchaux}, T. 2008, Astronomische Nachrichten, 329, 485

\bibitem[{{Appourchaux} {et~al.}(2002){Appourchaux}, {Andersen}, \&
  {Sekii}}]{apo02}
{Appourchaux}, T., {Andersen}, B., \& {Sekii}, T. 2002, in ESA Special
  Publication, Vol. 508, From Solar Min to Max: Half a Solar Cycle with SOHO,
  ed. A.~{Wilson}, 47--50

\bibitem[{{Appourchaux} {et~al.}(1998){Appourchaux}, {Gizon}, \&
  {Rabello-Soares}}]{apo98}
{Appourchaux}, T., {Gizon}, L., \& {Rabello-Soares}, M.-C. 1998, \aaps, 132,
  107

\bibitem[{{Appourchaux} {et~al.}(2008){Appourchaux}, {Michel}, {Auvergne},
  {Baglin}, {Toutain}, {Baudin}, {Benomar}, {Chaplin}, {Deheuvels}, {Samadi},
  {Verner}, {Boumier}, {Garc{\'{\i}}a}, {Mosser}, {Hulot}, {Ballot}, {Barban},
  {Elsworth}, {Jim{\'e}nez-Reyes}, {Kjeldsen}, {R{\'e}gulo}, \&
  {Roxburgh}}]{apo08}
{Appourchaux}, T., {Michel}, E., {Auvergne}, M., {et~al.} 2008, \aap, 488, 705

\bibitem[{{Baglin} {et~al.}(2006){Baglin}, {Michel}, {Auvergne}, {Catala},
  {Aerts}, {Alecian}, {Amado}, {Appourchaux}, {Ausseloos}, {Ballot}, {Barban},
  {Baudin}, {Berthomieu}, {Boumier}, {Bohm}, {Briquet}, {Charpinet}, {Cunha},
  {De Cat}, {Dupret}, {Fabregat}, {Floquet}, {Fremat}, {Garrido}, {Garcia},
  {Goupil}, {Handler}, {Hubert}, {Janot-Pacheco}, {Lambert}, {Lebreton},
  {Lignieres}, {Lochard}, {Martin-Ruiz}, {Mathias}, {Mazumdar}, {Mittermayer},
  {Montalban}, {Monteiro}, {Morel}, {Mosser}, {Moya}, {Neiner}, {Nghiem},
  {Noels}, {Oehlinger}, {Poretti}, {Provost}, {Renan de Medeiros}, {de Ridder},
  {Rieutord}, {Roca-Cortes}, {Roxburgh}, {Samadi}, {Scuflaire}, {Suarez},
  {Theado}, {Thoul}, {Toutain}, {Turck-Chieze}, {Uytterhoeven}, {Vauclair},
  {Vauclair}, {Weiss}, \& {Zwintz}}]{bag06}
{Baglin}, A., {Michel}, E., {Auvergne}, M., {et~al.} 2006, in ESA Special
  Publication, Vol. 1306, ESA Special Publication, 39--50

\bibitem[{{Bedding} \& {Kjeldsen}(2006)}]{bed06}
{Bedding}, T. \& {Kjeldsen}, H. 2006, in ESA Special Publication, Vol. 624,
  Proceedings of SOHO 18/GONG 2006/HELAS I, Beyond the spherical Sun

\bibitem[{{Benomar}(2008)}]{benomar}
{Benomar}, O. 2008, Communications in Asteroseismology, 157, 98

\bibitem[{{Brewer} {et~al.}(2007){Brewer}, {Bedding}, {Kjeldsen}, \&
  {Stello}}]{brewer}
{Brewer}, B.~J., {Bedding}, T.~R., {Kjeldsen}, H., \& {Stello}, D. 2007, \apj,
  654, 551

\bibitem[{{Chaplin} {et~al.}(1997){Chaplin}, {Elsworth}, {Howe}, {Isaak},
  {McLeod}, {Miller}, \& {New}}]{chaplin}
{Chaplin}, W.~J., {Elsworth}, Y., {Howe}, R., {et~al.} 1997, \mnras, 287, 51

\bibitem[{{Croll}(2006)}]{croll}
{Croll}, B. 2006, \pasp, 118, 1351

\bibitem[{{Fr{\"o}hlich} {et~al.}(1997){Fr{\"o}hlich}, {Crommelynck}, {Wehrli},
  {Anklin}, {Dewitte}, {Fichot}, {Finsterle}, {Jim{\'e}nez}, {Chevalier}, \&
  {Roth}}]{froehlich97b}
{Fr{\"o}hlich}, C., {Crommelynck}, D.~A., {Wehrli}, C., {et~al.} 1997,
  \solphys, 175, 267

\bibitem[{{Gizon} \& {Solanki}(2003)}]{gizon}
{Gizon}, L. \& {Solanki}, S.~K. 2003, \apj, 589, 1009

\bibitem[{{Gregory}(2005)}]{gregory}
{Gregory}, P.~C. 2005, {Bayesian Logical Data Analysis for the Physical
  Sciences: A Comparative Approach with `Mathematica' Support} (Bayesian
  Logical Data Analysis for the Physical Sciences: A Comparative Approach with
  `Mathematica' Support.~Edited by P.~C.~Gregory.~ISBN 0 521 84150 X
  (hardback); QA279.5.G74 2005 519.5'42 -- dc22; 200445930.~Published by
  Cambridge University Press, Cambridge, UK, 2005.)

\bibitem[{{Gregory}(2007)}]{gregory07}
{Gregory}, P.~C. 2007, \mnras, 374, 1321

\bibitem[{{Harvey}(1985)}]{har85}
{Harvey}, J. 1985, in ESA Special Publication, Vol. 235, Future Missions in
  Solar, Heliospheric \& Space Plasma Physics, ed. E.~{Rolfe} \& B.~{Battrick},
  199--+

\bibitem[{{Hastings}(1970)}]{hastings}
{Hastings}. 1970, Biometrika, 57, 97

\bibitem[{{Kallinger} {et~al.}(2009){Kallinger}, {Gruberbauer}, {Guenther},
  {Fossati}, \& {Weiss}}]{kall_neu}
{Kallinger}, T., {Gruberbauer}, M., {Guenther}, D.~B., {Fossati}, L., \&
  {Weiss}, W.~W. 2009, ArXiv e-prints, 0811.4686v1

\bibitem[{{Kallinger} {et~al.}(2008){Kallinger}, {Weiss}, {Barban}, {Baudin},
  {Carrier}, \& {De Ridder}}]{kal08}
{Kallinger}, T., {Weiss}, W.~W., {Barban}, C., {et~al.} 2008, A\&A, submitted

\bibitem[{{Michel} {et~al.}(2008){Michel}, {Samadi}, {Baudin}, {Barban},
  {Appourchaux}, \& {Auvergne}}]{mic08}
{Michel}, E., {Samadi}, R., {Baudin}, F., {et~al.} 2008, ArXiv e-prints
  (0809.1078)

\bibitem[{{Roberts} {et~al.}(1997){Roberts}, {Gelman}, \& {Gilks}}]{roberts}
{Roberts}, G.~O., {Gelman}, A., \& {Gilks}, W.~R. 1997, Ann. Appl. Prob., 7,
  110

\bibitem[{{Toutain} \& {Appourchaux}(1994)}]{toutain}
{Toutain}, T. \& {Appourchaux}, T. 1994, \aap, 289, 649

\end{thebibliography}

\end{document}